\documentclass[sigconf]{acmart}

\usepackage{tabularx}
\usepackage{booktabs}

\usepackage{graphicx}
\usepackage{tablefootnote}
\usepackage{tikz}
\usepackage{pgfplots}
\usepackage{todonotes}
\pgfplotsset{compat=1.18}

\usepackage{amsmath}

\usepackage{booktabs}
\usepackage{array}
\usepackage{tabularx}
\usepackage{multirow}
\usepackage{makecell}
\usepackage[table]{xcolor}
\usepackage{placeins} 

\newcolumntype{L}[1]{>{\raggedright\arraybackslash}p{#1}}

\definecolor{themegray}{gray}{0.78}
\definecolor{darkgray}{gray}{0.80}

\newcommand{\figendhack}{\vspace{-6pt}}
\newcommand{\tablecaptionhack}{\vspace{-6pt}}
\newcommand{\tableendhack}{\vspace{-6pt}}
\newcommand{\presectionhack}{\vspace{-4pt}}

\AtBeginDocument{%
}

\begin{document}

\title{Mapping the Emerging Curriculum for AI-Assisted Software Engineering via Syllabus Analysis}


\author{Francis Geng}
\affiliation{%
  \institution{University of California, San Diego}
  \city{La Jolla}
  \state{CA}
  \country{United States}}
\email{fgeng@ucsd.edu}

\author{Anshul Shah}
\affiliation{%
  \institution{University of California, San Diego}
  \city{La Jolla}
  \state{CA}
  \country{United States}}
\email{ayshah@ucsd.edu}

\author{Mia Chen}
\affiliation{%
  \institution{University of California, San Diego}
  \city{La Jolla}
  \state{CA}
  \country{United States}}
\email{mic048@ucsd.edu}

\author{Paul Denny}
\affiliation{%
  \institution{University of Auckland}
  \city{Auckland}
  \country{New Zealand}}
\email{p.denny@auckland.ac.nz}

\author{Juho Leinonen}
\affiliation{%
  \institution{Aalto University}
  \city{Espoo}
  \country{Finland}}
\email{juho.2.leinonen@aalto.fi}

\author{Bill Griswold}
\affiliation{%
  \institution{University of California, San Diego}
  \city{La Jolla}
  \state{CA}
  \country{United States}}
\email{bgriswold@ucsd.edu}

\author{Gerald Soosai Raj}
\affiliation{%
  \institution{University of California, San Diego}
  \city{La Jolla}
  \state{CA}
  \country{United States}}
\email{asoosairaj@ucsd.edu}

\author{Leo Porter}
\affiliation{%
  \institution{University of California, San Diego}
  \city{La Jolla}
  \state{CA}
  \country{United States}}
\email{leporter@ucsd.edu}

\renewcommand{\shortauthors}{Geng et al.}

\begin{abstract}

As Generative AI coding tools reshape professional software development,
universities have begun designing courses to prepare students for AI-assisted development workflows. By analyzing the syllabi of these courses, we can gather empirical evidence about
these courses, reveal how this emerging curricular area is being defined, and gain guidance for future curriculum design.We analyzed 23 publicly available syllabi and course materials of upper-division, credit-bearing courses that meet specific criteria, including explicitly
addressing Generative AI in software engineering. Through iterative qualitative coding, we characterized courses’ learning objectives, assessments, topics, and documented AI tools. Our analysis reveals commonalities and differences among these courses that allow researchers and educators to study and develop future courses.

\end{abstract}



\maketitle
\vspace{-0.2cm}
\section{Introduction}
Generative AI tools are changing software development practices. Chat-based assistants, code completion systems, and agentic development tools are increasingly used to support coding, debugging, testing, and documentation~\cite{terragni2025futureAIDrivenSE}. These developments raise curricular questions about what students should learn when AI tools are part of software development. Prior position and review papers argue that software engineering education should move beyond treating generative AI only as an academic-integrity concern and should instead examine how AI assistance changes the skills, workflows, and judgment expected of future software engineers~\cite{daun2023chatgptSEEducation,kirova2024llmSEEducation,sengul2024conversationalAISEEducation}.

These course-design questions are timely because established guidance from the ACM/IEEE-CS Joint Task Force for undergraduate software engineering programs dates back to SE2014~\cite{ardis2015se2014}.  This defines expected student outcomes, core knowledge, and representative curriculum structures, but predates widespread professional use of generative AI, and thus does not
address how AI-assisted development should be incorporated, what learning objectives should be emphasized, or how to assess student work involving AI tools.


Computing education research has begun to document the educational implications of generative AI. Broad work in computing education has synthesized opportunities, risks, and open questions raised by generative AI in programming and computing classrooms \cite{prather2023robots,denny2024computingGenerativeAI}. Within software engineering education, recent studies have examined student experiences with LLMs in project courses, AI use in team-based software engineering work, and the effects of tools such as GitHub Copilot in larger or unfamiliar codebases \cite{korpimies2024unrestrictedLLMs,rasnayaka2024llmSEProject,shah2025copilotLargeCodebases,baresi2025chatgptSEFiveCourses,salomon2025genaiWorkflowCollaboration,borghoff2025genaiStudentProjects,kharrufa2026llmTeamProjects,kim2026aiIntegrationSEEducation,roy2026benchmarkingAIToolsSEEducation}. Less is known, however, about how AI-assisted software engineering is currently being organized as a course-level curricular area. In particular, we lack empirical visibility into what such courses state as learning objectives, how they assess student work, what topics they include, and which AI tools they publicly document.

Prior syllabus- and course-material analyses have used public course artifacts to study curricular patterns, course policies, and instructional coverage in computing education, including generative AI policy statements and AI ethics instruction \cite{cunningham2024csedSyllabi,ali2025aipolicies,aljedaani2026aiEthicsSyllabusReview}. We use a similar approach because public examples of AI-assisted software engineering courses are still limited and vary in structure.

In this paper, we analyzed publicly available syllabi and course materials from 23 U.S. upper-division, credit-bearing courses that explicitly frame generative AI as part of software engineering and include graded coursework involving GenAI use for software engineering tasks. We study the following research questions:

\begin{itemize}
\item \textbf{RQ1:} What learning objectives and assessment methods are employed in AI-assisted software engineering courses?
\item \textbf{RQ2:} Which topics are commonly included under the umbrella of AI-assisted software engineering?
\item \textbf{RQ3:} What AI tools are documented in AI-assisted software engineering courses?
\end{itemize}

This paper contributes to computing education research by 
providing an empirical map of how AI-assisted software engineering is being taught and assessed in early, publicly available, course designs.
Rather than treating generative AI only as a classroom policy issue, tool-adoption question, or isolated instructional intervention, we study courses that position AI assistance as part of software engineering practice. By analyzing public course materials across institutions, we
identify emerging common practices 
around learning objectives, assessment structures, topic coverage, and tool choices. 
The findings we present here can support instructors and curriculum designers as they decide which AI-specific practices should supplement durable software engineering competencies, how AI-assisted work should be assessed, and how courses can remain relevant as tools change.
\vspace{-0.2cm}
\section{Background and Literature Review}





\subsection{Changes in Professional Software Development in the GenAI Era}

The professional practice of software engineering is changing as GenAI tools are incorporated across the entire software development lifecycle, from requirements elicitation to code review \cite{DORA2025AI}. A large-scale analysis of GenAI adoption among software companies reported that developers using GenAI completed 21\% more tasks and merged 98\% more pull requests~\cite{faros2025aiproductivityparadox}. The same report found that time dedicated to code review increased by 91\%, alongside a 154\% increase in pull request size~\cite{faros2025aiproductivityparadox}. As professional development becomes more \textit{agentic} \cite{alenezi2026rise}, with AI agents working with limited human intervention, the skills that software developers need in the GenAI era are shifting.

For instructors aiming to impart industry-relevant skills to their students, the rapidly-evolving software engineering industry represents a moving target for computing curricula. \citeauthor{kam2025professionals} studied AI-using professional developers by analyzing their work goals, tasks, and needed skills and knowledge~\cite{kam2025professionals}. Their findings suggest that AI-enhanced developers need both new competencies in using GenAI effectively and many durable software engineering competencies from the pre-GenAI era. These competencies span effective GenAI use, core software engineering, adjacent engineering skills, and adjacent non-engineering skills, including communication and collaboration~\cite{kam2025professionals}. In a similar vein, \citeauthor{alenezi2026rise} reviews studies of AI in software engineering practice and argues that software engineering education should address skills such as AI literacy, agent orchestration, AI interaction practices, and large-scale authentic projects built with AI~\cite{alenezi2026rise}.

\subsection{Changes in CS Courses in the GenAI Era}

Since 2023, several studies have highlighted educators' responses to the proliferation of GenAI tools and capabilities \cite{lau2023ban, sheard2024instructor, gordon2026taskforce, PratherBeyondHype, ali2025aipolicies, Zastudil2023Perspectives}. Two early studies of instructor perspectives on GenAI's impact on CS courses come from \citeauthor{lau2023ban} \cite{lau2023ban} and \citeauthor{Zastudil2023Perspectives} \cite{Zastudil2023Perspectives}. Together, these studies documented a range of instructor perspectives in 2023, soon after the emergence of ChatGPT as a capable programming assistant. Instructors noted the importance of student engagement and participation \cite{Zastudil2023Perspectives} while also expressing concerns about over-reliance and cheating \cite{lau2023ban}. Importantly, instructors shared ideas about assessments that are ``AI-proof,'' such as oral exams and process-based assessments \cite{lau2023ban}. In later work, \citeauthor{sheard2024instructor} asked instructors about changes to learning objectives, finding broad agreement that traditional objectives should be updated in response to GenAI~\cite{sheard2024instructor}. More recently, \citeauthor{PratherBeyondHype} conducted a survey of computing educators across the CS curriculum, revealing more concrete course-level changes \cite{PratherBeyondHype}. Their results showed that most instructors did not explicitly disallow AI usage and agreed that the skills needed to create software have changed \cite{PratherBeyondHype}. Several instructors mentioned adding assessments to ensure students understand the code they submit, such as a synchronous code review \cite{PratherBeyondHype}. The survey also revealed prevalent themes of teaching skills related to effective GenAI use, such as prompting, and shifting toward higher-stakes, proctored exams \cite{PratherBeyondHype}. In 2026, the \citeauthor{gordon2026taskforce} conducted another instructor survey to understand specific course-level changes that instructors have made due to GenAI \cite{gordon2026taskforce}. The survey showed a shift in assessment methods, toward more process-based assessments and oral exams rather than grading based on the correctness of students' submissions \cite{gordon2026taskforce}. Many instructors also reported weighting take-home assignments lower and weighting proctored exams and student engagement higher \cite{gordon2026taskforce}. 

While a steady line of work has surveyed instructors to understand changes to their classrooms, our present study directly analyzes course materials from publicly accessible software engineering course websites. The closest methodological precedents are \citeauthor{ali2025aipolicies}~\cite{ali2025aipolicies} and \citeauthor{aljedaani2026aiEthicsSyllabusReview}~\cite{aljedaani2026aiEthicsSyllabusReview}, who used syllabus reviews to examine GenAI course policies and AI ethics instruction. Our study uses a similar course-material approach, but focuses on courses where GenAI is not only a policy issue or ethics topic, but part of the stated learning objectives, assessments, topics, and tools of software engineering education.



\vspace{-0.2cm}
\section{Methods}
\subsection{Dataset Construction}

We identified potential course materials through structured Google searches conducted between Mar 10, 2026 and Mar 30, 2026 using the following queries:

\begin{quote}
\texttt{site:.edu ("genai" OR "generative AI" OR "LLM" OR "AI-assisted" OR "modern") ("software engineering" OR "software development") course}
\end{quote}

\begin{quote}
\texttt{site:.github.com ("genai" OR "generative AI" OR "LLM" OR "AI-assisted" OR "modern") ("software engineering" OR "software development") course}
\end{quote}

We used a search-first rather than institution-first strategy because AI-assisted software engineering courses are still emerging, relatively rare, and not always documented in stable university catalogs. Relevant materials appeared on instructor websites and GitHub repositories as well as university catalog or department pages, so the searches were used to identify candidate course materials before eligibility screening. 


Because these broad searches surfaced many pages outside the study scope, we used the search results as leads and applied two stages of screening. First, we screened for institutional eligibility. We retained candidates only if they were U.S.-based, credit-bearing, upper-division university courses. We excluded commercial training pages, non-credit or extended-studies offerings, online course platforms without evidence of university credit, duplicate pages, and other non-course materials. GitHub pages were retained only when the associated course could be verified through a university source, such as a syllabus, registrar listing, departmental page, or equivalent course-information page. This triage yielded 32 institutionally eligible course candidates.

Second, we manually screened these 32 candidates for content eligibility. To be included in the final dataset, a course had to satisfy all three content criteria below:

\begin{enumerate}
\item \textbf{Explicit GenAI framing:} The official title, description, syllabus, or course website explicitly referenced generative AI, LLMs, or AI coding tools in the context of software engineering or software development.
\item \textbf{Software engineering lifecycle (SDLC) coverage:} The course materials explicitly described at least two SDLC domains, such as planning, requirements analysis, design, coding, testing, deployment, or maintenance.
\item \textbf{Required GenAI for graded SE tasks:} The course included structured graded labs, assignments, or projects in which GenAI use was explicitly required for software engineering tasks.
\end{enumerate}

Table~\ref{tab:courses} presents the 23 courses that met the three content criteria out of the 32 institutionally eligible candidates.

\begin{table}[t]
\centering
\scriptsize
\caption{Final dataset of 23 public course materials analyzed. When available, linked course titles point to public syllabi, course websites, GitHub course pages, or public course-information pages used to identify and code each course.}
\label{tab:courses}

\begingroup
\setlength{\tabcolsep}{2pt}
\renewcommand{\arraystretch}{0.95}

\begin{tabularx}{\columnwidth}{@{}
  r
  >{\raggedright\arraybackslash}X
  >{\raggedright\arraybackslash}p{0.17\columnwidth}
  >{\raggedright\arraybackslash}p{0.22\columnwidth}
@{}}
\toprule
\textbf{\#} & \textbf{Course title} & \textbf{Term} & \textbf{Institution} \\
\midrule
1 & \href{https://www.cs.kzoo.edu/cs488/syllabus.html}{AI-Assisted Software Development} & Fall 2023 & Kalamazoo College \\

2 & \href{http://tianyi-zhang.github.io/files/CS59200_AI_Assisted_Software_Engineering_Syllabus.pdf}{AI-Assisted Software Engineering} & Spring 2024 & Purdue \\

3 & \href{https://hosting.cs.vt.edu/specialtopics/graduate/spring24/5914-atkinson.html}{AI Tools for Software Delivery} & Spring 2024 & Virginia Tech \\

4 & \href{https://lingming.cs.illinois.edu/courses/cs598lmz-s25.html}{Software Quality Assurance with Generative AI} & Spring 2025 & UIUC \\

5 & \href{https://www.mccormick.northwestern.edu/computer-science/academics/courses/descriptions/397-6.html}{Applied AI for Software Development} & Spring 2025 & Northwestern \\

6 & \href{https://ai-developer-tools.github.io/}{AI Tools for Software Development} & Fall 2025 & CMU \\

7 & Software Engineering with Generative AI\textsuperscript{*} & Fall 2025 & Harvard \\

8 & \href{https://github.com/gai4se/GAI4SE-Course?tab=readme-ov-file}{Generative AI for Software Engineering} & Fall 2025 & NC State \\

9 & \href{https://themodernsoftware.dev/}{The Modern Software Developer} & Fall 2025 & Stanford \\

10 & \href{https://www.cs.umd.edu/class/fall2025/cmsc398z/}{Effective Use of AI Coding Assistants and Agents} & Fall 2025 & UMD \\

11 & \href{https://danny.cs.colorado.edu/courses/csci7000-011_F25/index.html}{Generative AI-Powered Software Engineering} & Fall 2025 & CU Boulder \\

12 & \href{https://courses.cs.washington.edu/courses/cse490a2/25au/}{AI-Assisted Software Development} & Fall 2025 & UW \\

13 & \href{https://www.cs.virginia.edu/~se4ja/files/seLLMs.html}{Software Engineering and LLMs} & Fall 2025 & UVA \\

14 & \href{https://www.cs.cmu.edu/~113/}{Effective Coding with AI} & Spring 2026 & CMU \\

15 & \href{https://kelloggm.github.io/martinjkellogg.com/teaching/cs485-sp26/}{AI-Assisted Software Engineering} & Spring 2026 & NJIT \\

16 & \href{https://mpcs-courses.cs.uchicago.edu/2025-26/spring/courses/mpcs-51238-1}{Design, Build, Ship} & Spring 2026 & UChicago \\

17 & \href{https://github.com/lingming/software-agents}{Software Engineering with LLM Agents} & Spring 2026 & UIUC \\

18 & \href{https://ai4sd-s26-memphis.github.io/}{AI Tools for Software Development} & Spring 2026 & Univ. of Memphis \\

19 & \href{https://github.com/utah-cs3960-sp26/syllabus/blob/main/syllabus.md}{Vibe Coding} & Spring 2026 & Univ. of Utah \\

20 & \href{https://ucsd-cse-115-215.github.io/sp26/index.html}{Generative AI and Programming} & Spring 2026 & UCSD \\

21 & \href{https://fau.simplesyllabus.com/en-US/doc/w803j6asg/Spring-2026-1-Full-Term-COT-6930-001-Topics-in-Computer-Science?mode=view}{Generative AI Software Development Lifecycles} & Spring 2026 & Florida Atlantic Univ. \\

22 & \href{https://johnguerra.co/classes/aiCoding_spring_2026/}{AI-Assisted Software Engineering} & Spring 2026 & Northeastern \\

23 & \href{https://eecs498-aase.github.io/index.html}{Applied Agentic Software Engineering} & Fall 2026 & Univ. of Michigan \\
\bottomrule
\end{tabularx}
\figendhack
\endgroup

\end{table}

\begingroup
\renewcommand{\thefootnote}{\fnsymbol{footnote}}
\footnotetext[1]{Course website is no longer publicly available.}
\endgroup
\figendhack

\subsection{Data Analysis}
\enlargethispage{1\baselineskip}

For each included course, we collected publicly available syllabi, websites, schedules, and assignment pages when available. Because documentation varied across courses, we labeled each dimension using the materials most relevant to it: learning objectives and assessment methods were typically drawn from syllabi or official course descriptions, while topics and tools were more often identified from schedules, assignment pages, and course websites. We treated explicitly stated course outcomes, objectives, or goals as learning-objective entries, and schedule items, module headings, assignment topics, or named lecture topics as topic entries.

We used different coding procedures across dimensions. Learning objectives and topics required interpretive coding because course language varied substantially across institutions. Assessment methods and tools involved direct extraction and conservative normalization of explicit course elements, so these dimensions were labeled by one author using decision rules developed through discussion with the author group.

\textbf{Learning objectives.} Three authors labeled all 86 extracted learning objectives using open coding and negotiated agreement. Because labels continued to emerge during analysis, all three authors reviewed every objective rather than dividing the dataset in advance. The authors iteratively proposed, merged, and clarified labels with reference to the original course materials until agreement was reached, so we report negotiated agreement rather than a separate inter-coder reliability statistic. Final labels were then grouped into higher-level themes, following negotiated and team-based qualitative coding practices~\cite{garrison2006negotiated,cascio2019team}.

\textbf{Topics.} Three authors labeled 211 extracted topic entries. To develop the codebook, all three authors open-coded 93 topics across three rounds, comparing labels, resolving disagreements, and refining definitions after each round. Once the codebook stabilized and inter-coder reliability, measured using Fleiss' $\kappa$, reached 0.80, the remaining 118 topics were divided across the three coders~\cite{halpin2024inter}. Final topic labels were grouped into higher-level themes.

\textbf{Assessment methods and tools.} One author labeled assessment methods and tools using decision rules discussed with the author group. For assessment methods, the coder extracted grading categories, grade weights, and assignment descriptions from public course materials, then normalized explicit course-specific labels into common categories. Assessment features were labeled only when the relevant property was stated in the materials; for example, group work required an explicitly team-based graded component, and AI-required work required explicit GenAI use in a graded software engineering task. Ambiguous features were left unlabeled rather than inferred. For tools, the coder recorded only AI tools explicitly named in public course materials.
\vspace{-0.2cm}
\section{Results}
We organize the results by research question and report denominators in the relevant tables and figure. Because all included courses already met our GenAI, SDLC, and GenAI for graded tasks inclusion criteria, the results focus on variation within this scoped set rather than on the presence of those criteria.

\subsection{RQ1: Learning Objectives and Assessments}
\label{rq1}

\begin{table}
\centering
\small
\caption{Learning-objective themes and label prevalence across 12 courses with public learning objectives.\tablecaptionhack}
\label{tab:learning_objectives}

\begingroup
\setlength{\tabcolsep}{2pt}
\renewcommand{\arraystretch}{1.0}

\newcommand{\ASEThemeBarTwoCol}[1]{%
  \multicolumn{2}{@{}>{\columncolor[gray]{0.80}}p{\dimexpr\columnwidth\relax}@{}}{%
    \textbf{\textit{#1}}%
  }\\[-0.1em]
}


\newcommand{\ASECountBar}[1]{%
\begin{tikzpicture}[baseline=-0.5ex]
\fill[gray!18] (0,0) rectangle (2.6,0.17);
\fill[black] (0,0) rectangle ({2.6*(#1)/12},0.17);
\node[font=\scriptsize, fill=white, inner xsep=1pt, inner ysep=0pt] at (1.3,0.085) {#1};
\end{tikzpicture}%
}

\begin{tabular*}{\columnwidth}{@{}
  p{0.70\columnwidth}
  @{\extracolsep{\fill}}
  c
@{}}
\toprule
\textbf{Learning-objective label} & \textbf{Courses (n=12)} \\
\specialrule{0.4pt}{0pt}{0pt}

\ASEThemeBarTwoCol{Human-AI collaboration practices}
apply AI best practices & \ASECountBar{6} \\
manage AI context & \ASECountBar{2} \\

\ASEThemeBarTwoCol{Software and AI tool development}
build software with AI tools & \ASECountBar{5} \\
build AI-based SE tools & \ASECountBar{4} \\
build integrated software systems & \ASECountBar{2} \\
create technical artifacts with AI & \ASECountBar{1} \\

\ASEThemeBarTwoCol{Software and AI evaluation}
evaluate software and AI artifacts & \ASECountBar{5} \\
assess software design decisions & \ASECountBar{2} \\
test software with AI & \ASECountBar{2} \\

\ASEThemeBarTwoCol{Foundational knowledge and skills}
develop general SE competencies & \ASECountBar{4} \\
understand AI fundamentals & \ASECountBar{3} \\

\ASEThemeBarTwoCol{Research practices}
conduct and present research & \ASECountBar{4} \\

\ASEThemeBarTwoCol{Responsible AI use and judgment}
analyze AI limitations & \ASECountBar{2} \\
analyze AI ethics/responsibility & \ASECountBar{2} \\
analyze AI impact on SE & \ASECountBar{2} \\
evaluate when to trust AI & \ASECountBar{2} \\

\ASEThemeBarTwoCol{User-centered design}
apply user-centered design & \ASECountBar{2} \\
use AI in user-centered design & \ASECountBar{2} \\

\ASEThemeBarTwoCol{Software refinement practices}
practice iterative refinement & \ASECountBar{2} \\
debug software systematically & \ASECountBar{1} \\

\ASEThemeBarTwoCol{Learning and creative agency}
develop lifelong learning skills & \ASECountBar{2} \\
develop creative project ideas & \ASECountBar{1} \\
learn with AI tools & \ASECountBar{1} \\

\bottomrule
\end{tabular*}

\endgroup
\tableendhack
\tableendhack
\end{table}

\subsubsection{Learning objectives.}
Among the 12 courses with public learning objectives, the most common themes were human-AI collaboration, software and AI tool development, and software and AI evaluation (Table~\ref{tab:learning_objectives}). Together, these themes suggest that courses taught AI-assisted software engineering as more than code generation: students were often expected to collaborate with AI systems while building, refining, and evaluating software artifacts.

The most common objective, \emph{apply AI best practices}, captured goals related to using AI tools in principled, task-appropriate, iterative, and feedback-driven ways. A related but less frequent objective, \emph{manage AI context}, focused on structuring information provided to AI tools. These objectives point to human-AI collaboration as a recurring learning goal, although it was often described broadly rather than as a fully specified set of assessable competencies.

Other common objective themes focused on building software or AI-based tools and on evaluating software, AI artifacts, or design decisions. Some courses emphasized using AI tools to build software, while others emphasized building AI-based tools or systems for software engineering work. The software and AI evaluation theme captured objectives about assessing software and AI artifacts for qualities such as correctness, maintainability, and trustworthiness. Less frequent objective areas included responsible AI use, user-centered design, software refinement, and learning or creative agency. These patterns suggest that courses more consistently foregrounded building and evaluating AI-assisted software artifacts than explicitly emphasizing broader concerns such as responsibility, user experience, or creative agency in their stated objectives.

\subsubsection{Assessment categories.}

\begin{table}
\centering
\small
\caption{Assessment categories in 14 conventional percentage-based courses, ordered by the number of courses containing the category. Grade-weight summaries are calculated only among courses containing each category.\tablecaptionhack}
\label{tab:assessment_categories}
\setlength{\tabcolsep}{3pt}
\renewcommand{\arraystretch}{1.05}

\begin{tabular*}{\columnwidth}{@{}
  p{0.42\columnwidth}
  @{\extracolsep{\fill}}
  c
  c
  c
@{}}
\toprule
\textbf{Category} & \textbf{Courses} & \textbf{Median} & \textbf{Range} \\
\midrule
Participation / engagement & 14 & 17.5\% & 5--46\% \\
Project / capstone & 12 & 50\% & 30--85\% \\
Coding assignment & 9 & 20\% & 10--60\% \\
Review / reflection & 6 & 19\% & 10--40\% \\
Paper presentation & 4 & 20\% & 10--20\% \\
Quiz / exam & 4 & 15\% & 10--30\% \\
\bottomrule
\end{tabular*}

\vspace{0.15em}

\end{table}

Across the 14 conventional percentage-based courses, assessment was organized primarily around participation, project work, and practical coding activity (Table~\ref{tab:assessment_categories}). We normalized course-specific grading labels into shared categories, including project/capstone for larger projects or major deliverables, coding assignment for bounded homework or lab-style programming tasks, review/reflection for reflective or critical written work, and paper presentation for presentations of existing research.

The main pattern is that these courses assessed AI-assisted software engineering through ongoing practice rather than primarily through exams. Every course included participation or engagement, most included a project/capstone component, and many included coding assignments. Project/capstone work appeared frequently and often carried a large share of the grade, but its weight varied widely, suggesting a shared project-oriented tendency rather than a settled assessment model. Less frequent categories, including review/reflection, paper presentation, and quiz/exam components, show that some courses supplemented project work with written, research-oriented, or more traditional assessments, but these were not the dominant structure.


\subsubsection{Assessment features.}
\begin{table}
\centering
\small
\caption{Assessment features in 14 conventional percentage-based courses, ordered by the number of courses containing the feature. Grade-weight summaries are calculated only among courses containing each feature.\tablecaptionhack}
\label{tab:assessment_features}
\setlength{\tabcolsep}{4pt}
\renewcommand{\arraystretch}{1.05}

\begin{tabular*}{\columnwidth}{@{}
  p{0.42\columnwidth}
  @{\extracolsep{\fill}}
  c
  c
  c
@{}}
\toprule
\textbf{Feature} & \textbf{Courses} & \textbf{Median} & \textbf{Range} \\
\midrule
AI-required work & 14 & 70\% & 20--100\% \\
Programming work & 14 & 70\% & 20--95\% \\
In-class work & 14 & 31\% & 5--80\% \\
Group work & 12 & 55\% & 5--90\% \\
Proctored assessment & 3 & 20\% & 10--30\% \\
\bottomrule
\end{tabular*}

\vspace{0.15em}
\tableendhack

\end{table}

Assessment features describe the kinds of work embedded within graded components, rather than the assessment format itself (Table~\ref{tab:assessment_features}). Because required GenAI use in graded software engineering tasks was part of our inclusion criteria, the key result is the extent to which grades were attached to AI-required, programming, in-class, group, and proctored work.

The feature analysis reinforces that these courses assessed AI-assisted software engineering through practice-oriented work. AI-required and programming work carried high median grade weights, indicating that graded work commonly centered on producing or improving software artifacts with AI tools. In-class work appeared in every conventional percentage-based course, suggesting that instructors often retained structured synchronous activities alongside AI-supported out-of-class work. Group work was also common, but its grade weight ranged from minor to dominant, indicating variation in how strongly courses emphasized collaborative software engineering. Proctored assessment appeared in only a small minority of courses. Because features are non-mutually exclusive, these weights should not be summed across rows.

\subsection{RQ2: Topics Included in AI-Assisted Software Engineering}
\label{rq2}

\begin{table}
\centering
\small
\caption{Course-topic themes and label prevalence across 18 courses with public topic information.\tablecaptionhack}
\label{tab:topics}

\begingroup
\setlength{\tabcolsep}{2pt}
\renewcommand{\arraystretch}{1.0}

\newcommand{\ASETopicThemeBarTwoCol}[1]{%
  \multicolumn{2}{@{}>{\columncolor[gray]{0.80}}p{\dimexpr\columnwidth\relax}@{}}{%
    \textbf{\textit{#1}}%
  }\\[-0.1em]
}


\newcommand{\ASETopicCountBar}[1]{%
\begin{tikzpicture}[baseline=-0.5ex]
\fill[gray!18] (0,0) rectangle (2.6,0.17);
\fill[black] (0,0) rectangle ({2.6*(#1)/18},0.17);
\node[font=\scriptsize, fill=white, inner xsep=1pt, inner ysep=0pt] at (1.3,0.085) {#1};
\end{tikzpicture}%
}

\begin{tabular*}{\columnwidth}{@{}
  p{0.70\columnwidth}
  @{\extracolsep{\fill}}
  c
@{}}
\toprule
\textbf{Topic label} & \textbf{Courses (n=18)} \\
\specialrule{0.4pt}{0pt}{0pt}

\ASETopicThemeBarTwoCol{AI/ML foundations}
AI/ML fundamentals & \ASETopicCountBar{12} \\

\ASETopicThemeBarTwoCol{Evaluation and quality assurance}
testing & \ASETopicCountBar{10} \\
AI evaluation & \ASETopicCountBar{7} \\
software evaluation & \ASETopicCountBar{6} \\
AI-assisted software evaluation & \ASETopicCountBar{4} \\

\ASETopicThemeBarTwoCol{AI-assisted development practices}
agentic development & \ASETopicCountBar{10} \\
prompting & \ASETopicCountBar{9} \\
AI development environments & \ASETopicCountBar{8} \\
build AI-integrated software & \ASETopicCountBar{5} \\
vibe coding & \ASETopicCountBar{5} \\
tool-specific knowledge & \ASETopicCountBar{3} \\

\ASETopicThemeBarTwoCol{AI practice and impacts}
AI in practice & \ASETopicCountBar{8} \\
AI ethics and responsibility & \ASETopicCountBar{6} \\
AI trends and impacts & \ASETopicCountBar{6} \\

\ASETopicThemeBarTwoCol{Core software engineering lifecycle}
debugging & \ASETopicCountBar{7} \\
backend development & \ASETopicCountBar{6} \\
software design & \ASETopicCountBar{6} \\
deployment & \ASETopicCountBar{5} \\
frontend development & \ASETopicCountBar{5} \\
software documentation & \ASETopicCountBar{3} \\
brownfield development & \ASETopicCountBar{2} \\
SE fundamentals & \ASETopicCountBar{2} \\

\ASETopicThemeBarTwoCol{Development processes and planning}
user-centered development & \ASETopicCountBar{6} \\
project management & \ASETopicCountBar{4} \\
spec-based development & \ASETopicCountBar{3} \\
test-driven development & \ASETopicCountBar{3} \\

\bottomrule
\end{tabular*}

\endgroup
\tableendhack
\end{table}

The 18 courses with public topic information documented a broad but uneven topic mix (Table~\ref{tab:topics}). AI/ML fundamentals was the most common topic label, suggesting that many courses introduced foundational AI/ML concepts alongside applied software engineering rather than treating AI tools only as black-box coding assistants.

AI-assisted development practices were also prominent. Courses frequently documented agentic development, prompting, and AI development environments, while tool-specific knowledge appeared less often. This pattern suggests that many public topic lists emphasized transferable AI-assisted development practices rather than instruction tied to a single named tool.

Evaluation and quality assurance formed another major cluster. Testing, AI evaluation, software evaluation, and AI-assisted software evaluation were all visible across the dataset. This emphasis is important because AI-assisted software engineering raises questions not only about how students generate software artifacts, but also about how they judge, verify, and improve artifacts produced through AI-assisted development practices.

Public topic lists also included a range of software engineering lifecycle and process topics, including debugging, backend development, software design, deployment, frontend development, user-centered development, project management, specification, and test-driven development. These topics appeared unevenly, suggesting selective emphasis rather than uniform SDLC coverage. Courses most often documented topics related to implementation, testing, evaluation, design, deployment, and planning, while areas such as documentation, brownfield development, and general SE fundamentals were less visible in public materials.

Finally, several courses situated AI-assisted software engineering in broader professional and social contexts through topics such as AI in practice, AI ethics and responsibility, and AI trends and impacts. Taken together, the topic results suggest that courses commonly combined foundational AI/ML concepts, AI-assisted development practices, and selected software engineering activities. Rather than showing a single settled model of AI-assisted software engineering, the topic distribution shows how instructors are currently assembling this emerging curricular area from both AI-specific practices and established software engineering domains.

\enlargethispage{2\baselineskip}

\presectionhack
\subsection{RQ3: AI Tools Documented in Courses}
\label{rq3}
\FloatBarrier
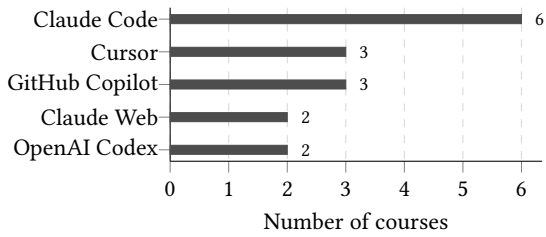
\begin{figure}[t]
\centering
\makebox[\columnwidth][l]{%
\begin{tikzpicture}
\begin{axis}[
    xbar,
    scale only axis,
    width=0.58\columnwidth,
    height=0.24\columnwidth,
    xmin=0,
    xmax=6.35,
    bar width=3.4pt,
    enlarge y limits={abs=0.35},
    clip=false,
    y dir=reverse,
    ytick=data,
    yticklabels={
        Claude Code,
        Cursor,
        GitHub Copilot,
        Claude Web,
        OpenAI Codex,
    },
    yticklabel style={
        align=right,
        anchor=east,
        inner xsep=0pt
    },
    xlabel={Number of courses},
    xtick={0,1,2,3,4,5,6},
    xmajorgrids=true,
    grid style={dashed, gray!30},
    axis x line*=bottom,
    axis y line*=left,
    tick align=outside,
    nodes near coords,
    nodes near coords align={horizontal},
    every node near coord/.append style={
        font=\footnotesize,
        anchor=west,
        xshift=2pt
    }
]
\addplot[
    fill=black!70,
    draw=black!70
]
coordinates {
    (6,0)
    (3,1)
    (3,2)
    (2,3)
    (2,4)
};
\end{axis}
\end{tikzpicture}%
}
\caption{Most common AI tools publicly listed by 9 courses. In addition, Aider, Amp, Anthropic API, ChatGPT, Gemini Code Assist, Google Antigravity, Ollama, OpenAI API, Replit, Willison's LLM CLI, and Warp were listed by one course. \tableendhack}
\label{ai_tool_frequencies}

\end{figure}
Of the 23 courses, 9 explicitly named one or more AI tools in public course materials; across these courses, we identified 16 distinct tools (Figure~\ref{ai_tool_frequencies}). Claude Code appeared most frequently, followed by Cursor and GitHub Copilot. The remaining named tools formed a long tail, with most appearing in only one course.

Among courses that publicly named tools, repeated references clustered around developer-facing coding tools rather than only general-purpose chat interfaces. Claude Code is described by Anthropic as an agent that reads codebases, edits files, and runs commands across the terminal, IDE, desktop app, and browser~\cite{anthropicClaudeCode}. Cursor describes itself as an AI coding agent for building software across desktop, CLI, web, and mobile surfaces~\cite{cursorProduct}. GitHub Copilot is similarly positioned as an AI coding assistant that works in the editor, command line, GitHub, and other development tools~\cite{githubCopilot}. This pattern suggests that, when public materials specified tools, they often pointed students toward tools embedded in development workflows rather than only standalone chat-based assistants.

At the same time, the full set of named tools was diverse. The long tail included general chat interfaces, model APIs, command-line tools, local-model tools, AI-enabled terminals, and development platforms. Thus, while a small cluster of developer-facing tools appeared repeatedly, the documented tool landscape remained fragmented, with no single tool dominating the public course materials.

These counts should be interpreted cautiously. Because we labeled only tools explicitly named in public materials, the results reflect documented tool references rather than all tools instructors introduced or students used. The fact that only 9 of 23 courses publicly named tools also suggests that tool choice may be underdocumented in public course artifacts, intentionally left flexible, or communicated through channels not captured by our dataset.

\vspace{-0.2cm}
\section{Limitations}

This study analyzes publicly documented course materials from U.S.-based courses, so the findings should be interpreted as patterns in public U.S. course artifacts rather than estimates of national or international prevalence. Courses behind login systems, with unpublished materials, non-indexed pages, or different terminology may be missing. Our inclusion criteria intentionally selected courses where generative AI was central to graded software engineering work, so the findings should not be generalized to courses where AI use is optional, peripheral, or documented only as a policy issue.

Public documentation varied across courses, and each analysis used the subset of courses with relevant available materials. Accordingly, our counts describe documented learning objectives, topics, assessments, and tools, not everything taught, assigned, or used in each course. Course-material analysis also cannot show how instructors enacted these courses, how students used AI tools in practice, or how effective the courses were for learning.
\vspace{-0.2cm}
\section{Discussion}
\enlargethispage{2\baselineskip}

\subsubsection*{Assessment and Individual Learning}

Assessment patterns highlight a tension between realistic AI-assisted workflows and assessing individual learning. Prior instructor-focused studies raised concerns about over-reliance and academic misconduct~\cite{lau2023ban,PratherBeyondHype,gordon2026taskforce} and documented responses such as process-focused assessment, oral or in-person checks, code-review or code-explanation activities, and proctored exams~\cite{PratherBeyondHype,gordon2026taskforce}. Our course-material analysis shows a different balance: courses relied heavily on project-based, programming-intensive, and collaborative work, while proctored assessment appeared in only a small minority of courses. This suggests that courses are emphasizing authentic AI-assisted practice while identifying mechanisms that make students' individual reasoning and process visible. Process logs, code reviews, demonstrations or oral defenses, individual reflections, and in-class checkpoints may help preserve project-based work while strengthening assessment of individual understanding.

\subsubsection*{Layering AI Practices onto Software Engineering Foundations}

In our data set, emerging AI-assisted software engineering courses appear to build on familiar software engineering education foundations. Many of the most visible elements are existing software engineering practices such as project work, programming, debugging, design and deployment. The newer curricular elements are the AI-mediated practices layered onto this foundation, such as prompting, agentic development, AI evaluation, managing context, and integrating AI into development workflows.

The documented course topics further show that AI-assisted software engineering is not framed only as tool use. Many courses covered AI/ML fundamentals, evaluation, testing, and selected software engineering lifecycle topics alongside AI-assisted development practices. This pattern aligns with prior arguments that students need more than prompt-level fluency: they also need conceptual and evaluative knowledge to guide AI use, inspect and validate outputs, recognize limitations, and decide when AI assistance is appropriate~\cite{sengul2024conversationalAISEEducation,alenezi2026rise}. Meanwhile, documentation, brownfield development, responsible AI, and user-centered design were less consistently visible. Future course designs may therefore need to clarify which underlying software engineering and AI concepts best complement tool practice.

\subsubsection*{Implications for Course and Curriculum Design}

For instructors, these findings identify common building blocks for course design: AI-assisted projects, programming-intensive assignments, collaborative development, evaluation and testing, and reflection on AI use. For departments, they show that introducing such courses requires curricular decisions about scope, prerequisites, assessment, and placement, not simply decisions about which tools to teach. For computing education researchers and curriculum designers, the results offer a baseline for studying how AI-assisted software engineering education evolves as tools, professional practices, and curricular expectations continue to change.

\presectionhack
\section{Conclusion}

This study shows that emerging AI-assisted software engineering courses commonly center project or capstone work, require AI use, substantial programming, and collaboration while positioning AI use within realistic software development workflows rather than isolated tool exercises. Meanwhile, the courses do not yet reflect a single curricular model. Many emphasize transferable practices, including prompting, agentic development, evaluation, testing, and workflow integration, but differ in tool choices, topics, and assessment structures.
Our findings suggest that AI-assisted software engineering remains a design space rather than a standardized curriculum. By documenting design patterns across current public courses, this study provides a baseline for comparing future course designs and tracking how AI-assisted software engineering education develops.

\begin{acks}
This work is supported by the NSF Graduate Research Fellowship Program and NSF Award \#2417374, and by the Google Award for Inclusion Research Program. Any opinion, findings, and conclusions or recommendations expressed in this material are those of the authors(s) and do not necessarily reflect the views of the NSF.
\end{acks}

\clearpage
\bibliographystyle{ACM-Reference-Format}
\bibliography{refs}

\end{document}